%% file: main.tex
\documentclass[a4paper,11pt]{article}
\pdfoutput=1 % if your are submitting a pdflatex (i.e. if you have
\usepackage{jheppub} % for details on the use of the package, please
\usepackage[T1]{fontenc} % if needed

\input{symbols.tex}

\title{\boldmath Testing the Higgs Boson Coupling to Gluons}

\author[a]{U.~Langenegger,}
\author[a]{M.~Spira,}
\author[a,b]{I.~Strebel}

\affiliation[a]{Paul Scherrer Institute, CH-5232 Villigen PSI, Switzerland}
\affiliation[b]{ETH Zurich, CH-8093 Zurich, Switzerland}

\emailAdd{urs.langenegger@psi.ch}
\emailAdd{michael.spira@psi.ch}
\emailAdd{strebeli@phys.ethz.ch}

\abstract{ We study the possibility to separate in gluon fusion
  loop-induced Higgs boson production from point-like production. The
  Higgs boson is reconstructed in the $\hgg$ final state at very large
  transverse momentum. Using the Higgs boson yields (normalized to the
  overall rate) and the shape of the Higgs boson \pt\ distribution the
  two hypotheses can be separated with 2 standard deviations with an
  integrated luminosity of about $500\invfb$. The largest experimental
  uncertainty affecting this estimate is the background event
  yield. The theoretical uncertainties from missing top mass effects
  are large, but can be decreased with dedicated calculations. }

\begin{document} 
\begin{flushright}
PSI--PR--15--07
\end{flushright}
\maketitle
\flushbottom

\newlength\fwidth

\input{introduction.tex}
\input{simulation.tex}

\input{analysis.tex}
\input{results.tex}
\input{conclusions.tex}
\input{acknowledgements.tex}


%\bibliographystyle{phaip}
%\bibliographystyle{lesHouches}
%\bibliography{main}
%\end{document}

\end{document}

%% file: introduction.tex
\section{Introduction}
\label{s:introduction}

\begin{figure}[h]
\hspace{3cm} \includegraphics[width=1.20\textwidth]{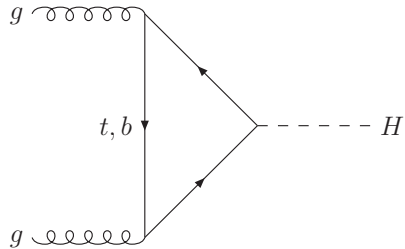} \\[-20cm]
\caption{\label{fg:ggF} \it Gluon fusion $gg\to H$ at leading order
mediated by top and bottom triangle loops.}
\end{figure}

\begin{figure}[h]
\includegraphics[width=1.20\textwidth]{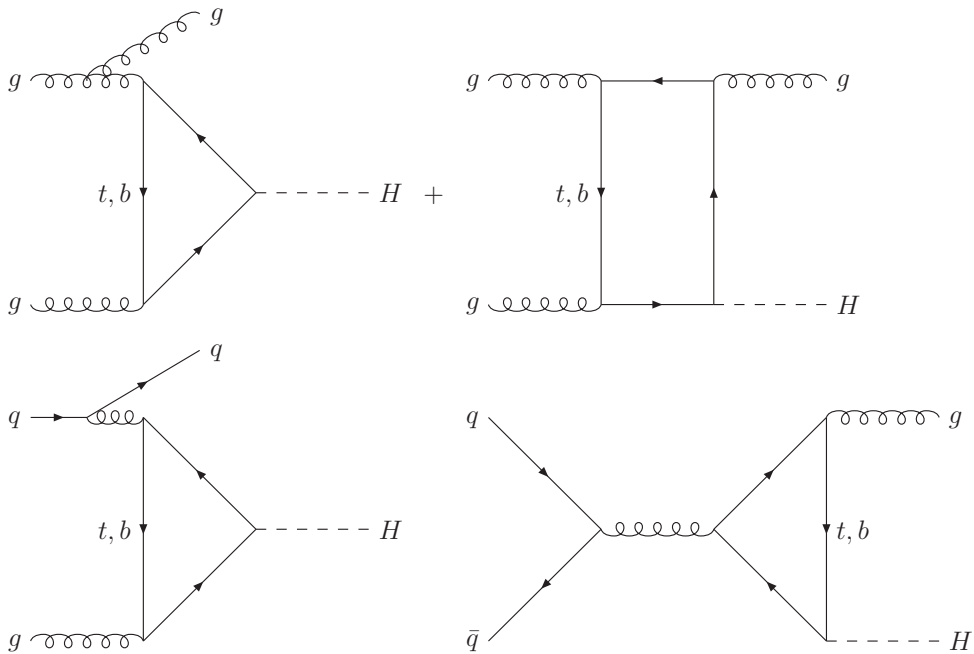} \\[-14cm]
\caption{\label{fg:ggFpt} \it Generic diagrams for Higgs production in
association with a jet via gluon fusion at leading order mediated by top
and bottom triangle loops generated by $gg, gq, q\bar q$ initial states.}
\end{figure}

Since the discovery of a Higgs-like state at the LHC
\cite{higgsdiscovery} huge efforts have been made to study its
properties in detail \cite{hcoup} in order to test consistency
with the predictions of a Standard Model (SM) Higgs particle
\cite{higgs}. Its couplings to vector bosons and fermions agree with SM
expectations within errors. The present state of the analyses at the LHC
underline the ${\cal CP}$-even and spinless nature of the discovered
particle. The dominant Higgs boson production mechanism
within the SM is the loop-induced gluon-fusion process $gg\to H$ that
is mediated by top and to a lesser extent bottom triangle loops as
exemplified in Fig.~\ref{fg:ggF}.  Although the total inclusive
production rate agrees with the SM prediction within uncertainties, it
is not clear that this production process is indeed loop-induced. For
an experimental proof of the loop nature of the gluon-fusion process the
top mass effects of the formfactors describing the Higgs coupling to
gluons have to be measured. A point-like
Higgs coupling to gluons is defined in terms of the effective Lagrangian
\begin{equation}
{\cal L}_{eff} = c_g \frac{\alpha_s}{\pi} G^{a\mu\nu} G^a_{\mu\nu}
\frac{H}{v}
\end{equation}
with $\alpha_s$ denoting the strong coupling constant, $G^{a\mu\nu}$ the
gluon field strength tensor, $H$ the Higgs field and $v$ the electroweak
vacuum expectation value.  The Wilson coefficient $c_g$ is adjusted to
reproduce the measured inclusive cross section. In order to distinguish
between a point-like Higgs coupling to gluons and the loop-induced type
a large energy scale has to be inserted into the loop. At LO the gluon
fusion process proceeds at the fixed scale given by the Higgs mass. The
first distribution that gives rise to a variable energy scale inside the
loops is the transverse momentum distribution of the Higgs particle that
is generated primarily by the additional radiation of a gluon at LO (see
Fig.~\ref{fg:ggFpt}). The comparison of the $p_T$ distribution at small
and large values of $p_T$ allows for a distinction between top mass
effects due to the loop contributions and a top mass independent
point-like Higgs coupling to gluons \cite{hinchliffe}. First studies of
the prospects at the LHC to resolve the Higgs coupling to gluons have
been made recently \cite{resolveloop}.  These studies, however, are all
theory-based while a rigorous experimental study is missing so far. The
purpose of this letter is to describe and present the results of our
analysis of the $p_T$ spectrum of the Higgs particle with a detailed
inclusion of experimental effects.  In this way our study goes
significantly beyond those of Refs.~\cite{resolveloop}.

The NLO corrections to the $p_T$-distribution are only known in the
limit of a heavy top quark \cite{ptnlo} supplemented by subleading terms
in the inverse top mass at NLO\,\cite{ptnlomt}. As for the inclusive
cross section the QCD corrections are large and positive.  Recently the
NNLO QCD corrections to the $p_T$ distribution have been derived in the
heavy top limit yielding a further moderate increase of $\sim 30\%$
\cite{ptnnlo}, thus corroborating a reliable perturbative behaviour.

Since the pure LO and NLO results diverge for $p_T\to 0$, the small
$p_T$ region requires a soft gluon resummation for a reliable
prediction. This resummation has been performed systematically for the
top quark loops in Ref.\,\cite{ptresum}, neglecting finite top mass
effects at NLO.  Soft gluon effects factorize, so that the top mass
effects at small $p_T$ are well approximated by the LO mass dependence
for small Higgs
masses\,\cite{Alwall:2011cy,Bagnaschi:2011tu,Mantler:2012bj}. Since the
top-loop contribution dominates the cross section for the SM Higgs
boson, the only limiting factor of the NLO+NNLL result as implemented in
{\tt HqT} or {\tt HRes}\,\cite{grazzinicodes} is thus the heavy-top
approximation of the NLO corrections which affects the whole $p_T$ range
for large Higgs masses and the large $p_T$ region in particular for all
Higgs masses. It has been shown that the subleading NLO terms in the
inverse top mass affect the $p_T$ distribution by less than 10\% for
$p_T \lsim 300$ GeV, if the full LO mass dependence is taken into
account \cite{ptnlomt}.

In addition to the leading top-quark contribution the bottom quark loops
amount to about $-6\%$ in the inclusive cross section \cite{gghnlo}.
However, at leading order the bottom quark contribution is only sizable
for small $p_T$, while for larger $p_T$ values it can safely be
neglected.  Recently, bottom quark contributions have been included in
the predictions for the resummed
$p_T$-distributions\,\cite{Bagnaschi:2011tu,Alwall:2011cy,%
Grazzini:2013mca,Mantler:2012bj,Banfi:2013eda,Harlander:2014uea,%
Bagnaschi:2015qta}. However, soft-gluon-resummation and bottom-quark
effects do not play a role in our analysis that just concentrates on the
shape at large $p_T$ values, where the resummed predictions coincide
with the fixed-order results.

%% file: simulation.tex
\section{Simulation setup}
\label{sec:simulation}

The focus of this study is on Higgs boson decays to two photons, as
this final state provides a reasonably large signal-to-background
$(S/B)$ ratio combined with a fair Higgs boson transverse momentum
(\pt) resolution. The alternative decay $H\to ZZ^{(*)}\to4\ell\, (\ell
= e,\mu)$ would offer superior $S/B$ and \pt\ resolution, but suffers
from even smaller statistics due to the very small effective branching
fraction. Signal (background) event samples corresponding to
proton-proton collisions at $\sqrt{s} = 14\tev$ are produced with NLO
(LO) event generators.  Subsequently these data are passed through a
parametrized detector simulation.

The signal events \hgg\ are generated with \mcatnlo\
4.10~\cite{Frixione:2002ik, mcatnlo} with a dynamic scale of $\mu =
\frac12\,\sqrt{m_H^2 + \pt^2}$, where $m_H = 125\gev$ is the Higgs
boson mass. For the loop-induced process, the exact top and bottom
quark mass dependence is taken into account following the two-step
procedure advocated in Ref.~\cite{Grazzini:2013mca}.  The events are
hadronized with \hpp~\cite{Bahr:2008pv} using the cluster model
of~\cite{Webber:1983if}; the shower evolution is performed with the
algorithm described in~\cite{Marchesini:1983bm}. For systematic
studies, the QCD renormalization $\mu_R$ and factorization $\mu_F$
scales are varied according to the recipe provided
in~\cite{LHCHXSWGGGF}, i.e., from 0.5 to 2.0 times the nominal value
$\mu$, excluding $\mu_R/\mu_F = 1/4$ and $4$. This variation not only
changes the total production cross section, but also affects the shape
of the Higgs boson high-\pt\ distribution. The
MSTW08~\cite{Martin:2009iq} set of parton distribution functions (PDF)
is used as default, while for systematic studies, NNPDF
2.3~\cite{Ball:2012cx} and CT10~\cite{Lai:2010vv} are considered. The
total gluon fusion cross section for Higgs boson production is
normalized to $\sigma = 49.5\pb$ with the uncertainties mentioned in
Ref.~\cite{LHCHXSWGGGF}. Eventually, the total gluon fusion cross
section will be measured with a substantially better experimental
precision.  The branching fraction for the considered final state is
$\cbf(\hgg) = 2.28\times10^{-3}$, with a relative uncertainty of 5\%
\cite{Denner:2011mq} which has been obtained from combining results of
\hdecay \cite{hdecay} and \prophecy4f \cite{prophecy4f}.

Using \powheg~\cite{powheg} to simulate the two signal Higgs boson
scenarios instead of \mcatnlo\ does not change the conclusion of this
paper in a significant way. 

The diphoton background events are generated with \sherpa\
V1.4.0~\cite{sherpa14} with leading-order MSTW08 PDFs. In addition
to the two photons, one or two jets are allowed in the process. Three
requirements are applied at the generator level: (1) the generated
photon energy is required to be $E>30\gev$, (2) the invariant mass of
the two photons must be in the range $50 < m_{\gamma\gamma} < 200\gev$
and (3) the diphoton \pt\ has to fulfill $\pt > 130\gev$. In this
kinematic region, the total cross section determined by \sherpa\ is
$\sigma_{\mathrm{tot}} = 1.6\pb$. We assume an uncertainty of 40\% on
this cross section in the high-\pt\ region.

The diphoton \pt\ distribution has been measured by ATLAS and CMS in
$pp$ collisions at $\sqrt{s} = 7\tev$. ATLAS showed in
Ref.~\cite{atlas:diphoton} that for $\pt> 300\gev$ the diphoton sample
is dominated by two real photons and that contributions from one or
two jets misidentified as photons are roughly at the 10\% level. The
high-\pt\ data are very well described in shape by the \sherpa\ (and
\pythia) MC simulations. For the rate, the ATLAS collaboration scaled
the expectations of \sherpa~1.3.1 up by 20\%.  The CMS
collaboration~\cite{cms:diphoton} found very good agreement of
\sherpa~1.4.0 (the same version is used for the present analysis) with
the differential and integrated diphoton cross sections, albeit
restricted to somewhat lower diphoton \pt.

The detector response simulation is based on
\delphes3~\cite{delphes}. The default card for the CMS detector,
included in the \delphes3 distribution, is used. One modification is
applied to the card: the cone for the calculation of photon isolation
is reduced from $\Delta R = \sqrt{\Delta\eta^2 + \Delta\phi^2} = 0.5$
to $\Delta R = 0.1$. The original setting leads to a noticeable
inefficiency for Higgs bosons measurements at very large \pt, as the
two photons are reconstructed close to each other. The performance of
the ATLAS detector would likely be comparable, given the very similar
performance of CMS and ATLAS in the Higgs boson measurements to date.

A significant limitation at the high-luminosity LHC (HL-LHC) will be
the pile-up, currently estimated at around 140 overlapping events per
bunch crossing. At present there is no complete understanding on the
exact details of photon measurement and identification performance for
that LHC and detector configuration. The current `particle-flow'
approach (also implemented in \delphes3) is likely to be replaced by
more powerful algorithms, which will allow for much better mitigation
of pile-up effects than the present photon reconstruction. For the
time being, we do not include pile-up in the detector simulation and
account for this with a large selection efficiency uncertainty. It
should be stressed that in this analysis no jet information is used,
the only ingredients are high-\pt\ photons.

We assume furthermore that events with \hgg\ candidates at high \pt\
can be triggered with a negligible inefficiency.

%% file: analysis.tex
\section{Analysis}
\label{s:analysis}

The purpose of this analysis is an estimate of the expected
sensitivity to discriminate between the loop-induced and a point-like
Higgs boson coupling. For this purpose, Higgs boson samples
corresponding to the two hypotheses combined with diphoton background
samples are studied. The invariant mass distribution provides the
primary means to separate the Higgs boson signal from the diphoton
background. As the expected number of Higgs bosons in the high-\pt\
region differs between the two hypotheses, the invariant mass
distribution also provides a powerful handle to distinguish between
the two hypotheses. In addition the \pt\ distribution, where the two
hypotheses lead to different shapes, provides additional separation
power as illustrated in Fig.~\ref{plot1} (left).

% -- differential pT cross section
\setlength\fwidth{0.49\textwidth}
\begin{figure}[htbp]
  \begin{centering}
    \includegraphics[width=\fwidth]{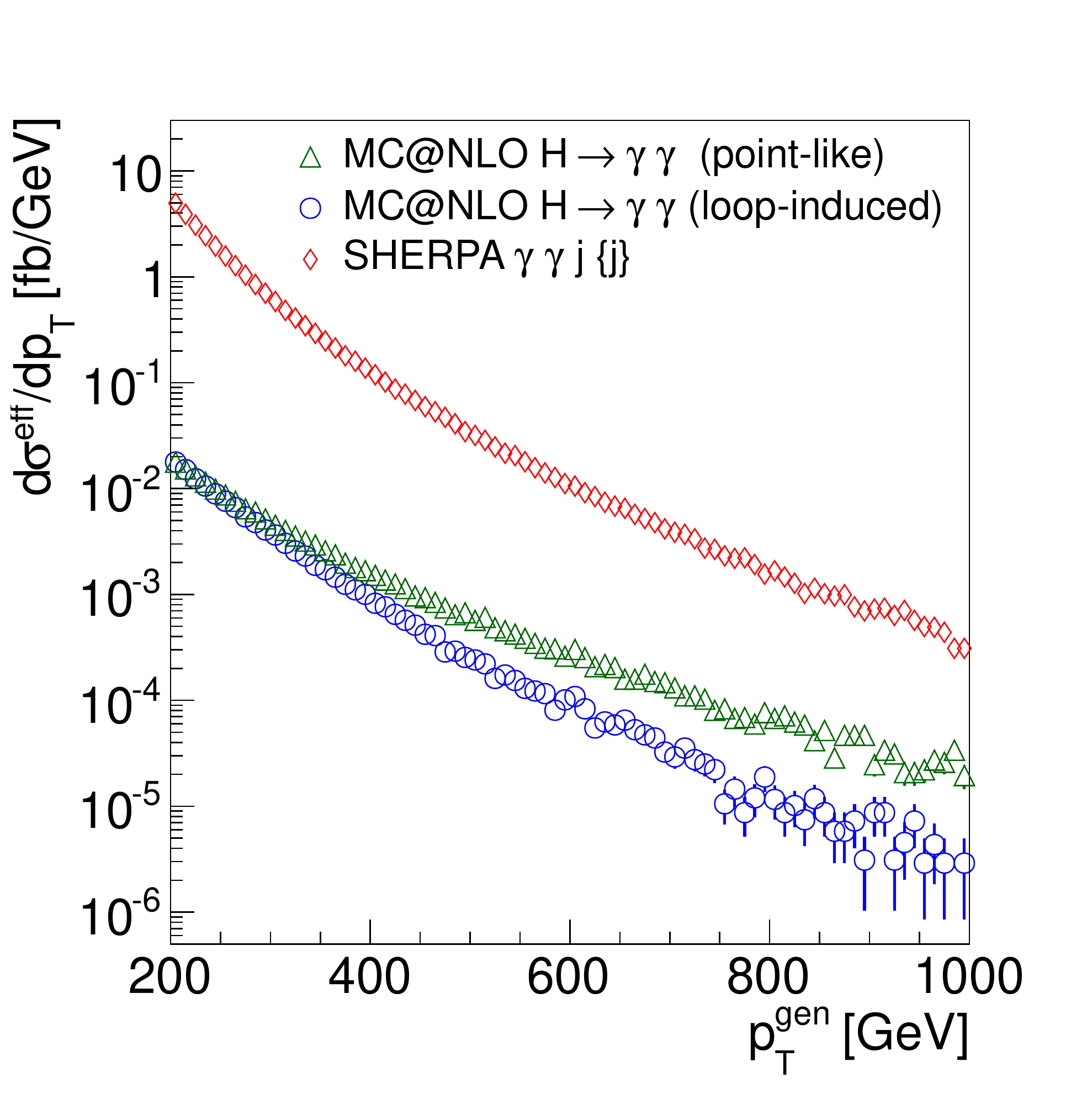}
    \includegraphics[width=\fwidth]{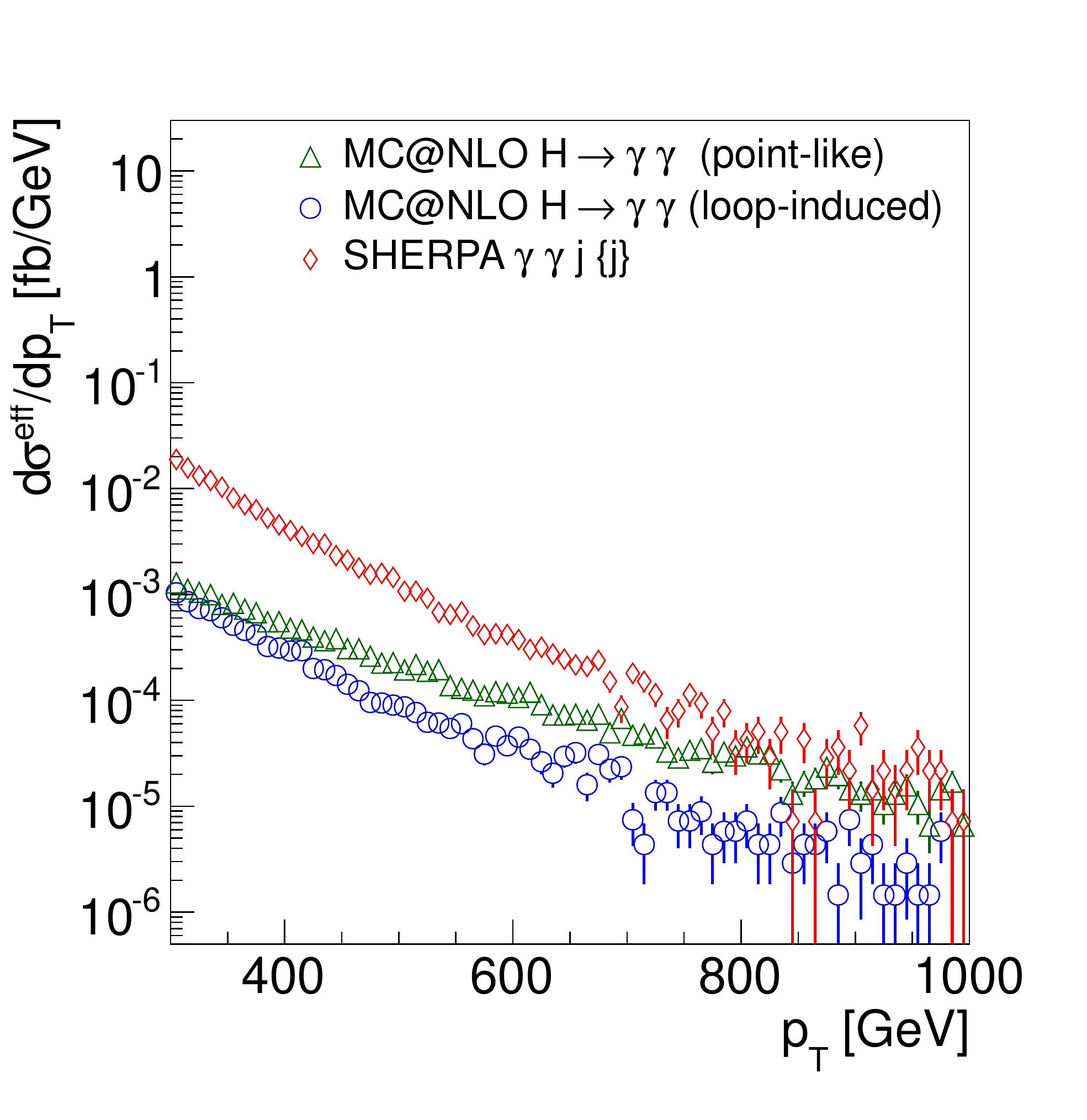}
    \caption{Differential production cross section as generated (left)
      and reconstructed after all analysis requirements (right). The
      effective cross section includes $\cbf(\hgg)$ for the Higgs
      boson cross sections (left and right),  and the analysis
      efficiency for all cross sections (right). The error bars show
      the statistical uncertainty only.}
    \label{plot1}
  \end{centering}
\end{figure}
% \clearpage

The data analysis uses reconstructed objects from the \delphes3
detector simulation. No attempt is made to unfold the true Higgs boson
\pt\ spectrum from the `reconstructed' spectrum. Rather the
`reconstructed' \pt-spectra from Higgs boson signal hypotheses and
diphoton background are used to fit the combined `data' spectra.

% ----------------------------------------------------------------------
\subsection{Event Reconstruction}
\label{ss:reco}
Reconstructed and identified photons from \delphes3 are subject to
additional requirements. The isolation requirement of each photon is
further tightened: the maximum energy flow in the form of tracks or
energy clusters (except for the other photon forming the Higgs boson
candidate) allowed within a cone of $\Delta R < 0.3$ around the photon
momentum is required to have $\pt < 0.5\gev$. The efficiency of this
requirement is 52\% for signal events and 17\% for background events,
nearly independent of the diphoton \pt\ requirement. This requirement
is quite strict, but not inconsistent with present
capabilities~\cite{Khachatryan:2015iwa} using area-based energy
subtraction procedures~\cite{Cacciari:2007fd}. As mentioned above, an
improved pileup treatment will be required for the successful usage of
the photon isolation variable at the HL-LHC. 

The two photons with the largest \pt\ are combined to form Higgs boson
candidates. The diphoton invariant mass must fulfill $70 < \mgg <
180\gev$.  The leading (subleading) photon has to fulfill $\ET >
80(50)\gev$ and for both photons $|\eta| < 2.5$ is required. No jet
information is used in this analysis, even though implicitly there
will be jets as we require for the diphoton $\pt > 300\gev$ (this will
be referred to as high-\pt\ region).  For this kinematic region the
signal (background) selection efficiency is 32.3\% (3.4\%). The values
for the single and diphoton \pt\ requirements were optimized by
maximizing the expected hypothesis discrimination sensitivity.

In Fig.~\ref{plot1} (right) the effective differential cross section
for signal and background after the event selection is shown. The
background contribution is found to be very large even after the above
selection. The average expected diphoton invariant mass and \pt\
distributions for an integrated luminosity of 1000\invfb\ are shown in
Fig.~\ref{plot2}, where signal and background have been combined (in
contrast to Fig.~\ref{plot1}, where signal and background are shown
separately).  The Higgs boson invariant mass resolution depends
strongly on the diphoton \pt\ and is limited by the angular resolution
in the kinematic range of this analysis.

% -- event yields normalized to 1000/fb
\setlength\fwidth{0.49\textwidth}
\begin{figure}[htbp]
  \begin{centering}
    \includegraphics[width=\fwidth]{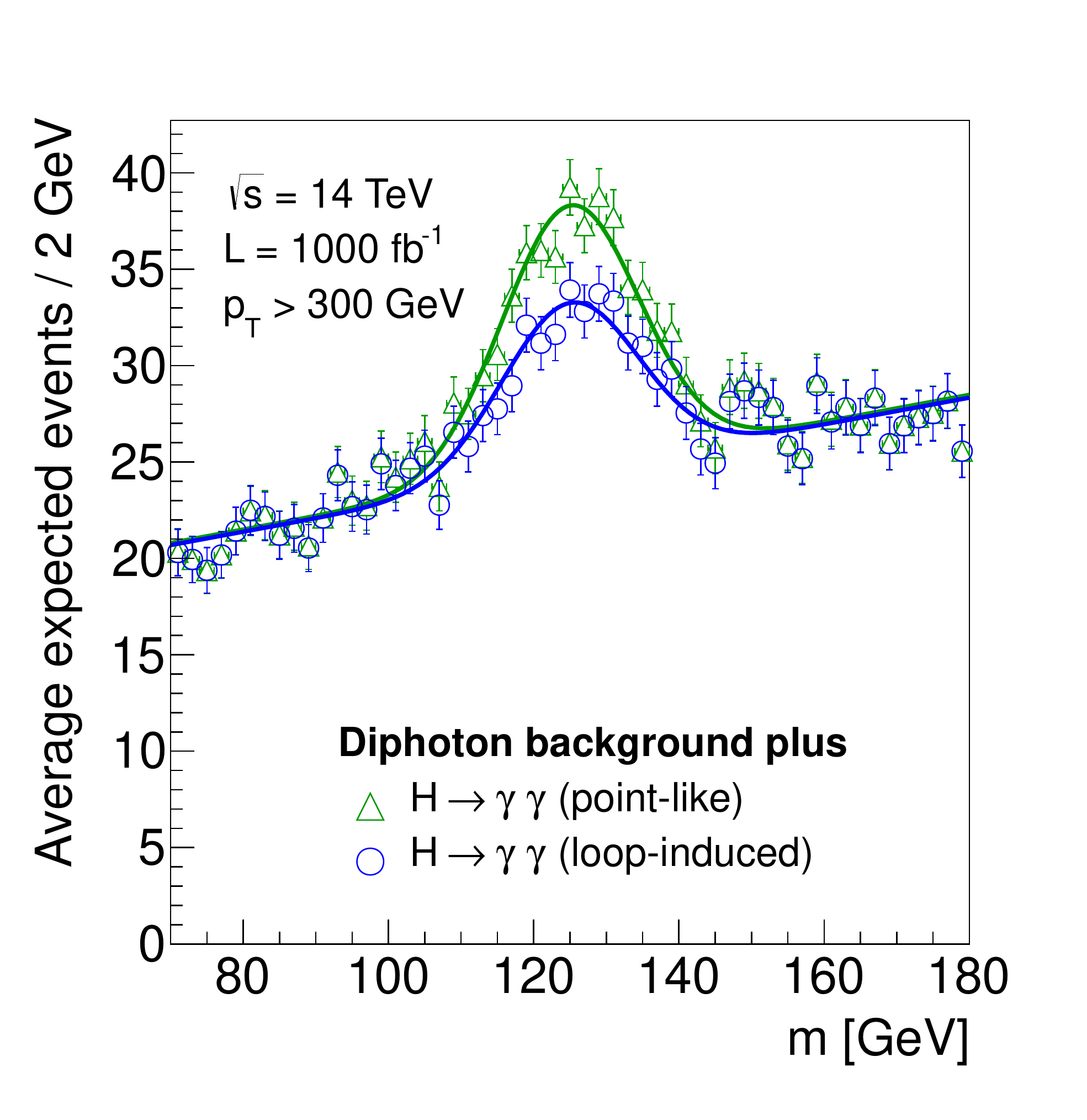}
    \includegraphics[width=\fwidth]{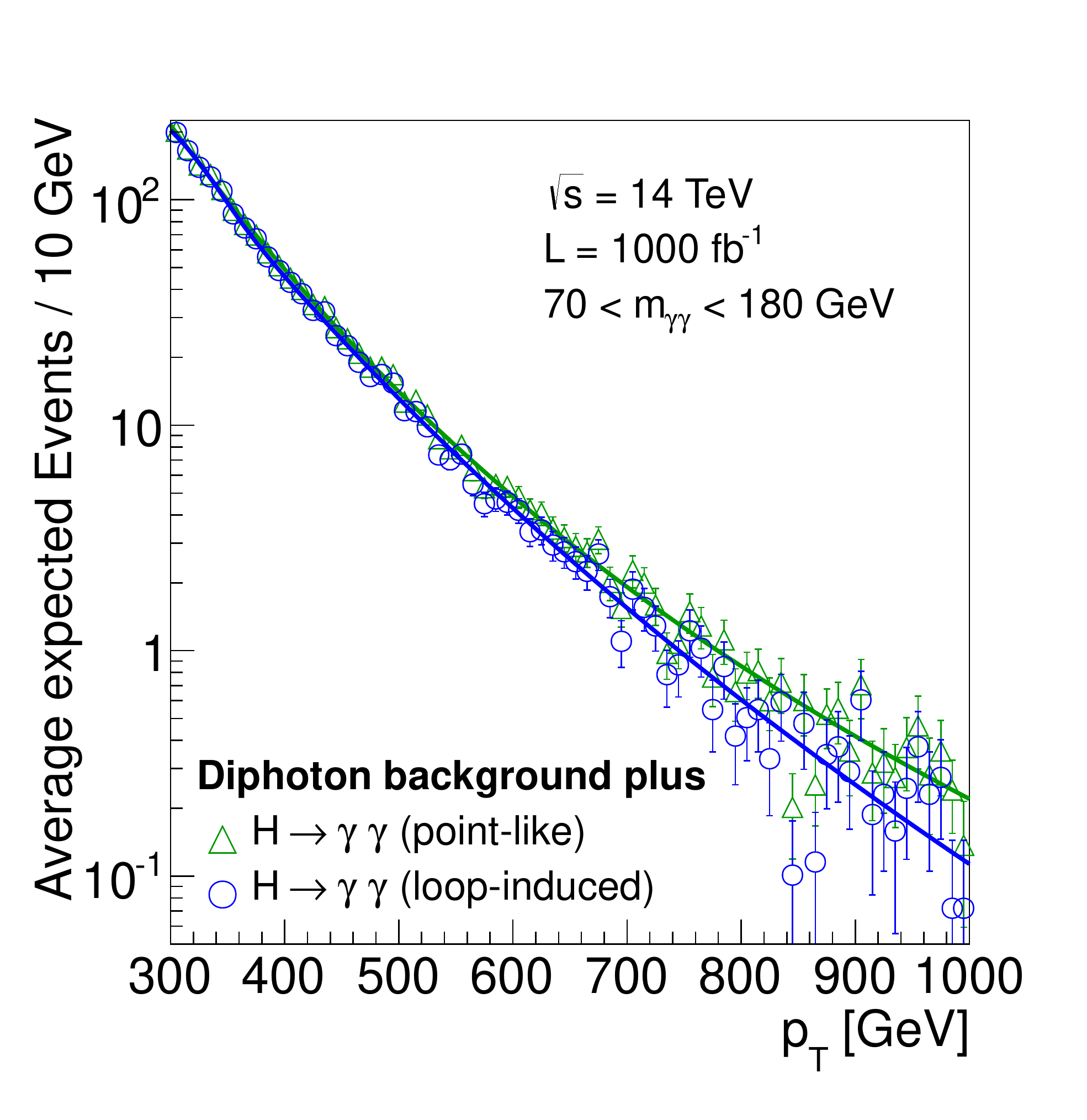}
    \caption{Average expected event yields vs diphoton invariant mass
      (left) and diphoton \pt\ (right), normalized to 1000\invfb. The
      histograms contain the background plus the point-like Higgs
      production (open triangles) and the background plus the
      loop-induced Higgs production (open circles). The background
      component is identical for both hypotheses. The error bars show
      the statistical uncertainty only.}
    \label{plot2}
  \end{centering}
\end{figure}
% \clearpage

The invariant mass distribution for the signal is well described by a
single Gaussian with standard deviation $\sigma = 9.4\gev$ in the
high-\pt\ region. The background is parametrized with a first degree
polynomial. The rising background invariant mass shape arises from the
selection criteria applied, in particular the very high-\pt\
requirement and the photon isolation criteria.  In the high-\pt\
region, the shape of the signal and background \pt\ distributions can
be described with a log normal distribution.  The free parameters of
the fit are: the Higgs boson yield, the background yield, two
parameters ($\mu$ and $k$) each for the signal and background log
normal distributions, and the slope of the polynomial describing the
background mass distribution. The position and width of the Higgs
boson peak are fixed.

% ----------------------------------------------------------------------
\subsection{Sensitivity Analysis}
\label{ss:sensitivity}

The sensitivity study is based on a two-dimensional extended unbinned
maximum likelihood fit to the diphoton invariant mass and \pt\
distributions using the probability density functions described above.
The $H_0$ hypothesis corresponds to the scenario of loop-induced Higgs
boson production plus the diphoton background. The $H_1$ hypothesis
describes the point-like Higgs boson production together with the
diphoton background.

Toy data are generated for the two hypotheses, where the event numbers
in the toy datasets depend on the integrated luminosity studied. In
Fig.~\ref{plot3} an example toy data set is shown. In the high-\pt\
region and for a luminosity of $\clu = 1000\invfb$ the expected
background yield is $N_{B} = 1360$, while the signal components for
the respective hypotheses are $N_S^{H_0} = 86$ and $N_S^{H_1} = 150$
(see below for a discussion of the systematic uncertainties).  From
the negative log-likelihood ratio for the two hypotheses, a test
statistic $q = -2 \ln(\clu_{H_{1}}/\clu_{H_{0}})$ is constructed.
%The two
%hypotheses are distinguished through the Higgs boson yield and the
%shape of the Higgs boson \pt\ distribution. 
The background parameters
(normalization and the shape of the mass distribution and the shape of
the \pt\ distribution) are treated as nuisance parameters and profiled
over. The expected separation between $H_0$ and $H_1$ is quantified as
follows~\cite{Cousins:2005pq,Gao:2010qx,Chen:2013waa}: We determine a midpoint
$\tilde{q}$ between the medians of the two test statistic
distributions such that ${\cal P}(q \ge \tilde{q}|H_1) = {\cal P}(q
\le \tilde{q}|H_0)$. This tail probability is converted into a
significance $\tilde{Z}$ using the one-sided Gaussian tail
convention. For the separation power we quote $Z = 2\tilde{Z}$, as the
midpoint $\tilde{q}$ is half-way between the medians of the two test
statistic distributions.
Using 10\,000 toy data sets a relative sensitivity uncertainty of about
1\% is achieved. To study systematic uncertainties, different toy data
sets are generated corresponding to the uncertainty under study.

% -- event yields normalized to 1000/fb
\setlength\fwidth{0.49\textwidth}
\begin{figure}[htbp]
  \begin{centering}
    \includegraphics[width=\fwidth]{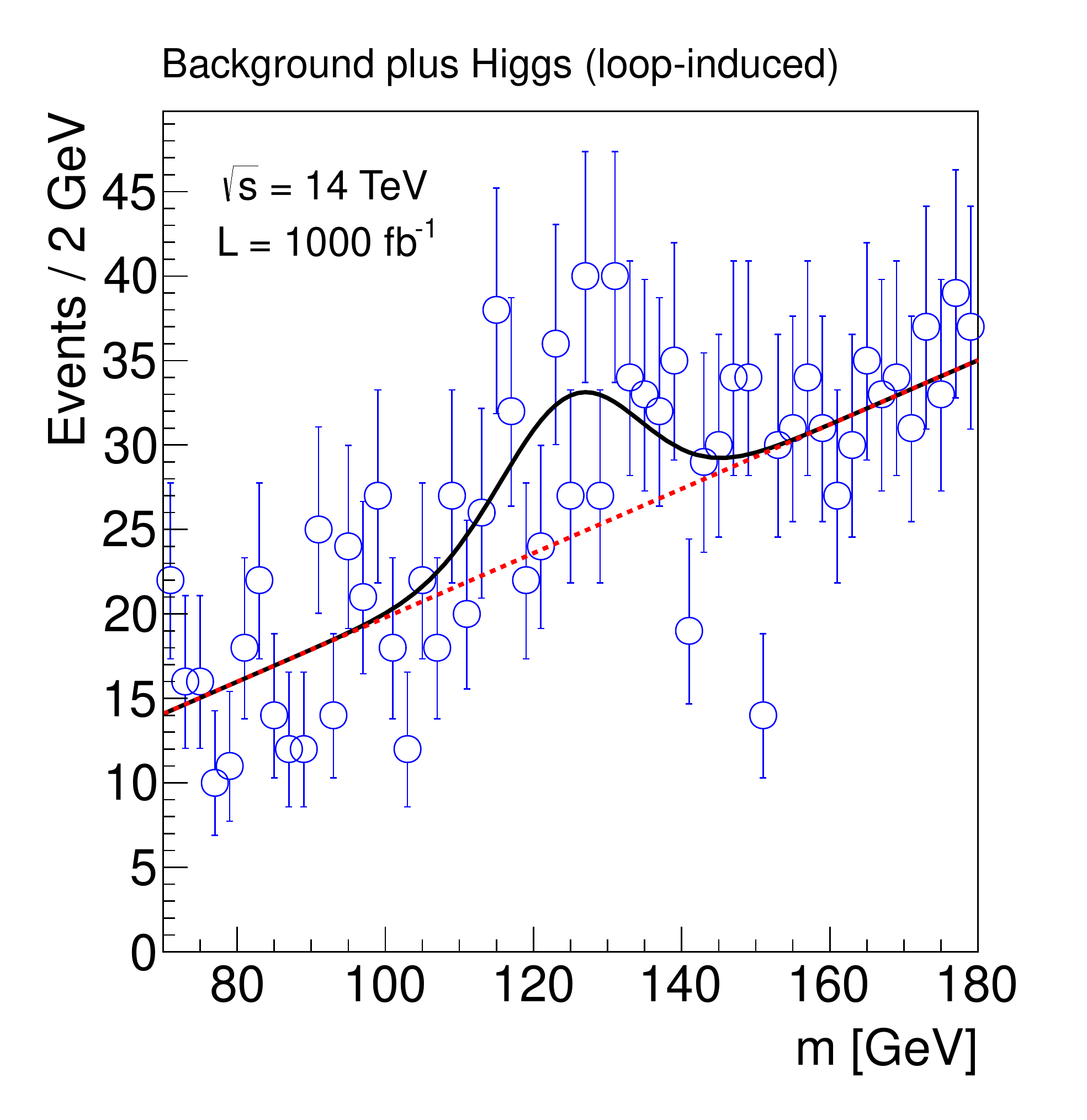}
    \includegraphics[width=\fwidth]{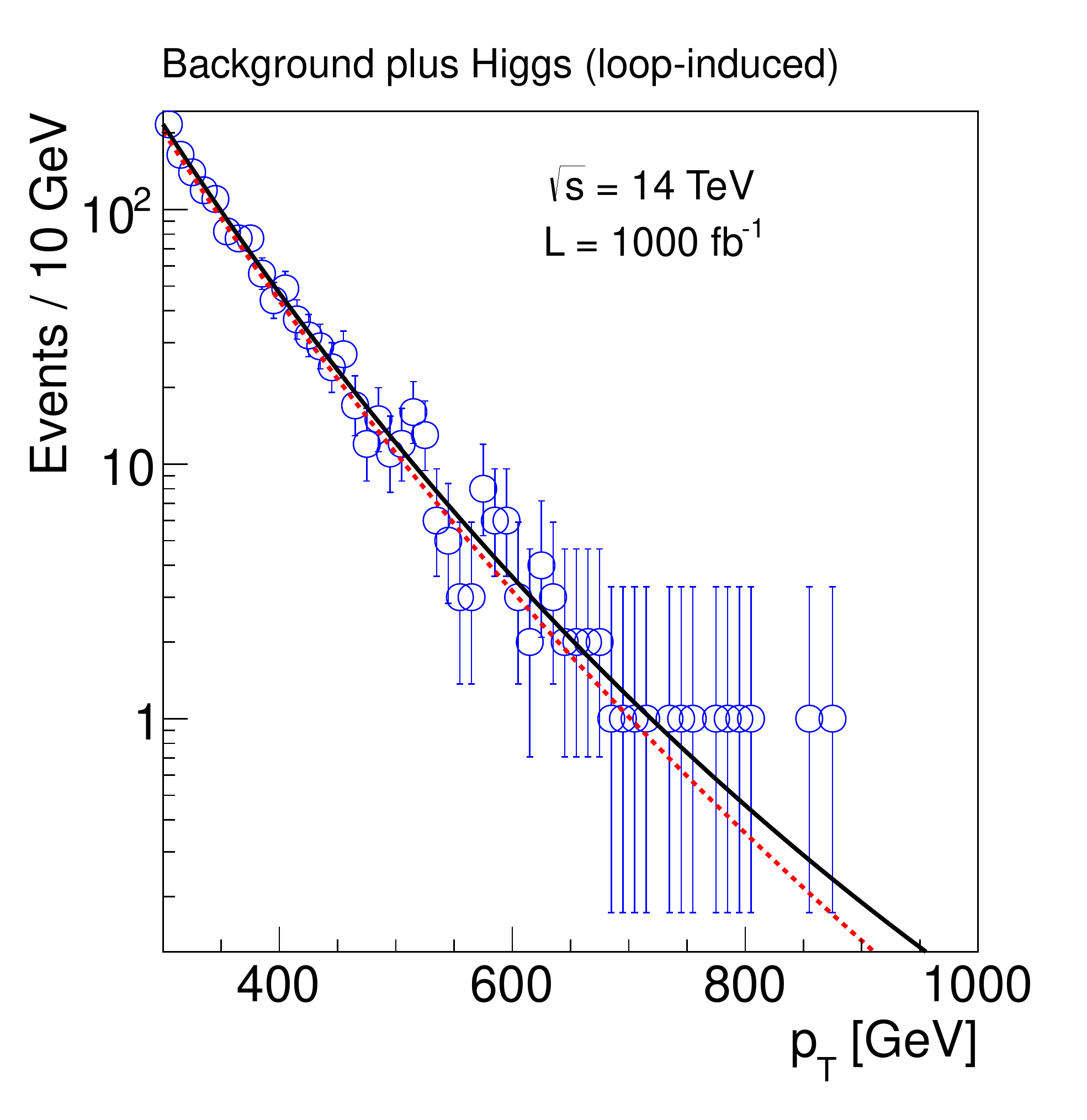}
    \includegraphics[width=\fwidth]{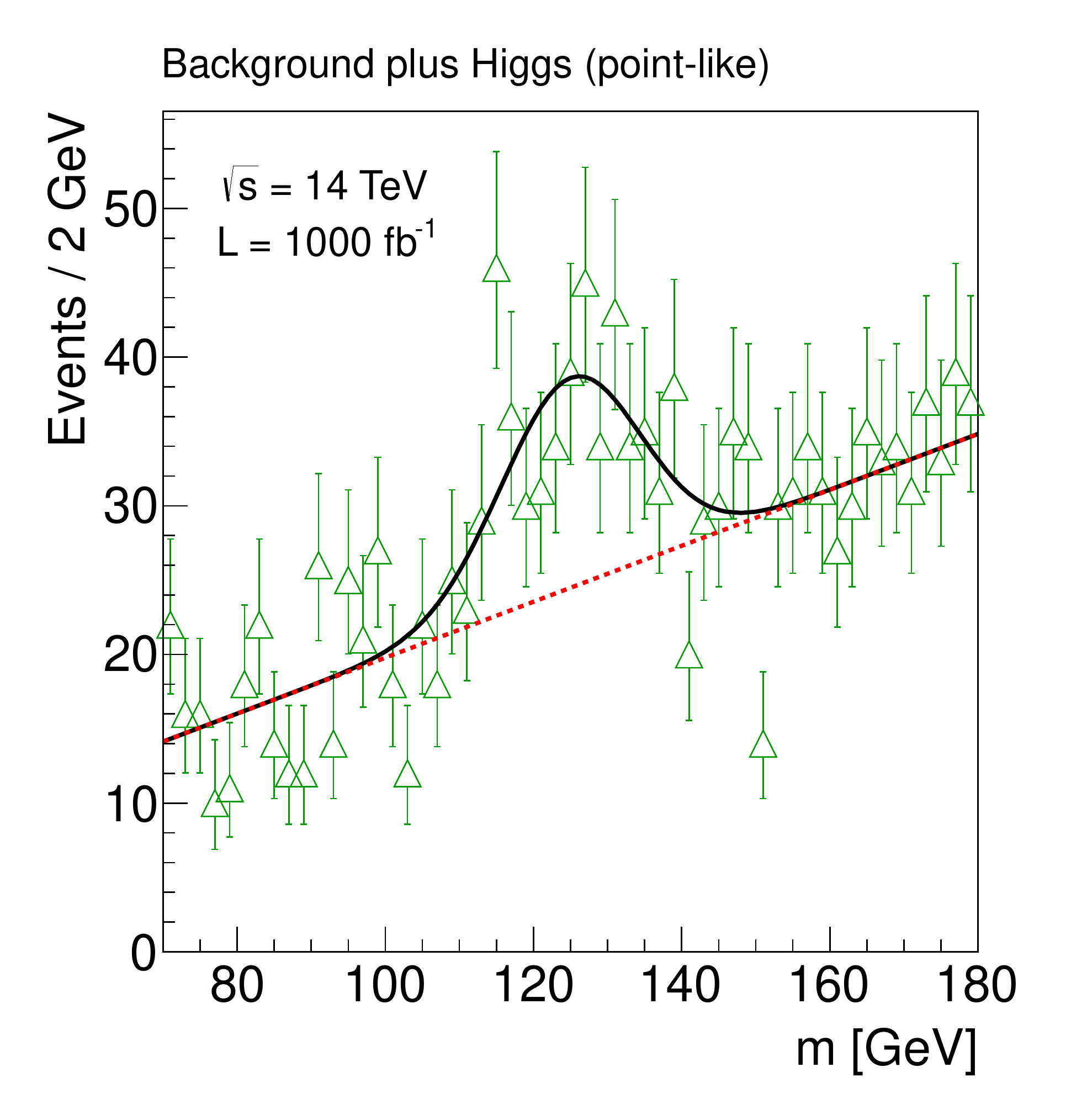}
    \includegraphics[width=\fwidth]{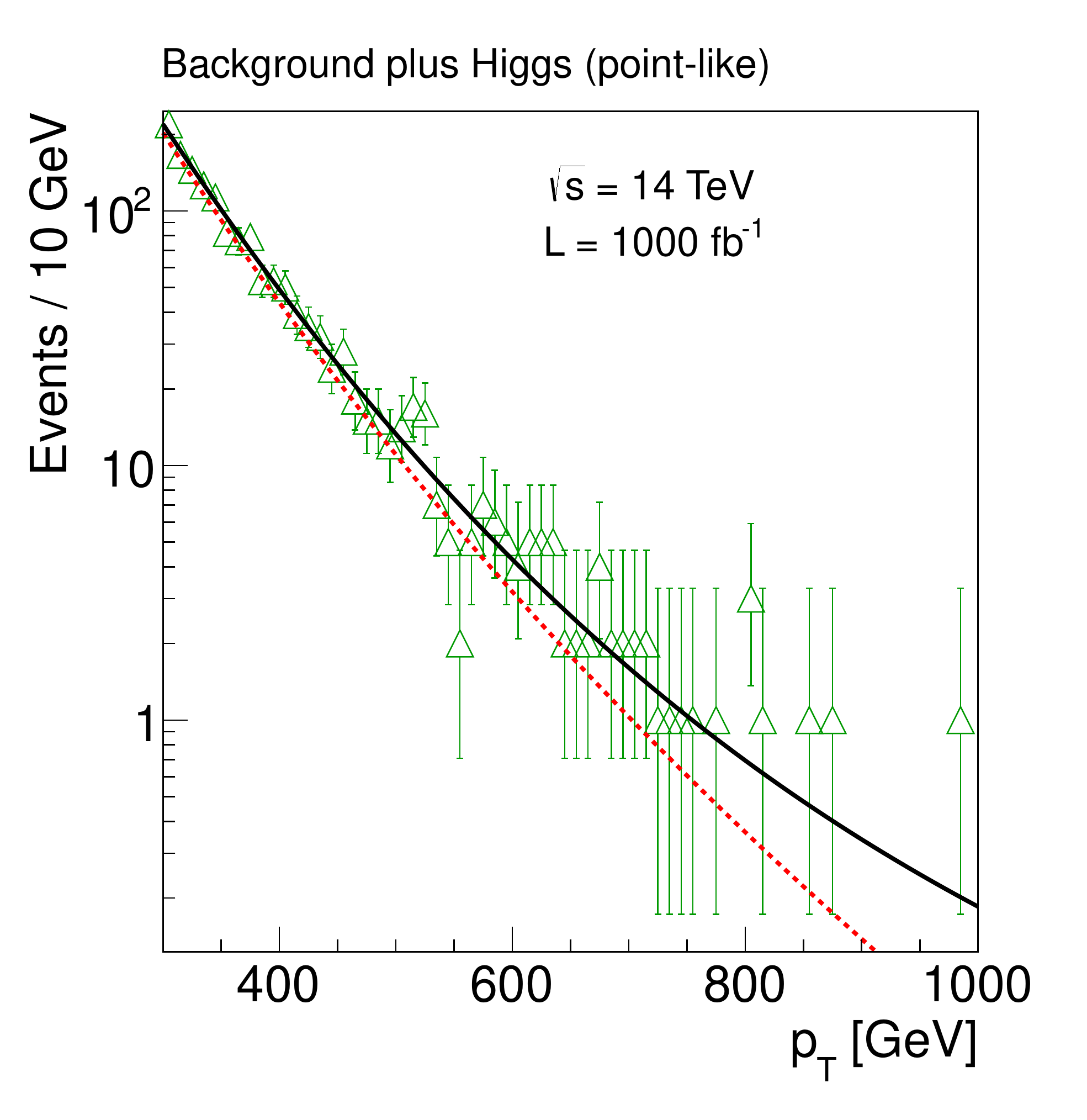}
    \caption{Example toy data set with overlayed projections of the
      extended unbinned maximum likelihood fits for the diphoton
      invariant mass (left) and the diphoton \pt\ (right). The $H_0$
      hypothesis (loop-induced Higgs boson production) is shown in the
      top row, the $H_1$ hypothesis (point-like Higgs boson
      production) is shown in the bottom row. The background component
      is identical for both hypotheses. The (black) solid curves shows
      the signal+background fit, the (red) dashed line shows the
      background component. The error bars show the statistical
      uncertainty only.}
    \label{plot3}
  \end{centering}
\end{figure}
% \clearpage

%% file: results.tex
\section{Results}
\label{s:results}

In Fig.~\ref{result-lumi} the result of the sensitivity study is
shown. The central expectation corresponds to the expected signal and
background numbers as discussed above. Systematic uncertainties on the
expected sensitivity arise from various sources.

Experimental uncertainties have a significant impact as this
sensitivity study ventures into a pile-up regime where adequate
algorithms are still to be developed. For the signal efficiency we
assume a relative uncertainty of $\pm20\%$. As no diphoton measurement
exists for the kinematic range considered in this analysis, and since
no jet to photon misidentification is considered, an uncertainty of
$\pm40\%$ is used for the background yield.  The sampling uncertainty
from the limited statistics in the toy data sets is minor for the
sensitivity regime found here.

Theoretical uncertainties have a larger impact on the expected
sensitivity.  The scale uncertainties from $\mu_R$ and $\mu_F$ affect
both the event yields for the two hypothesis and (to a smaller extent)
the shape of the \pt\ distributions.  The missing top mass effects are
parametrized with
$$\frac{\Delta x}{x} = \left(\frac{\pt-40\gev}{100\gev}\right)^2 \times 1.5\%,$$
\noindent where $x = d\sigma/d\pt$. This parametrization has been
adjusted to the known systematic top mass effects obtained in
Ref.~\cite{ptnlomt}. Uncertainties from a variation of the PDFs or the
top quark mass are found to be negligible.

% -- pT cross section distributions
\setlength\fwidth{0.49\textwidth}
\begin{figure}[htbp]
  \begin{centering}
    \includegraphics[width=\fwidth]{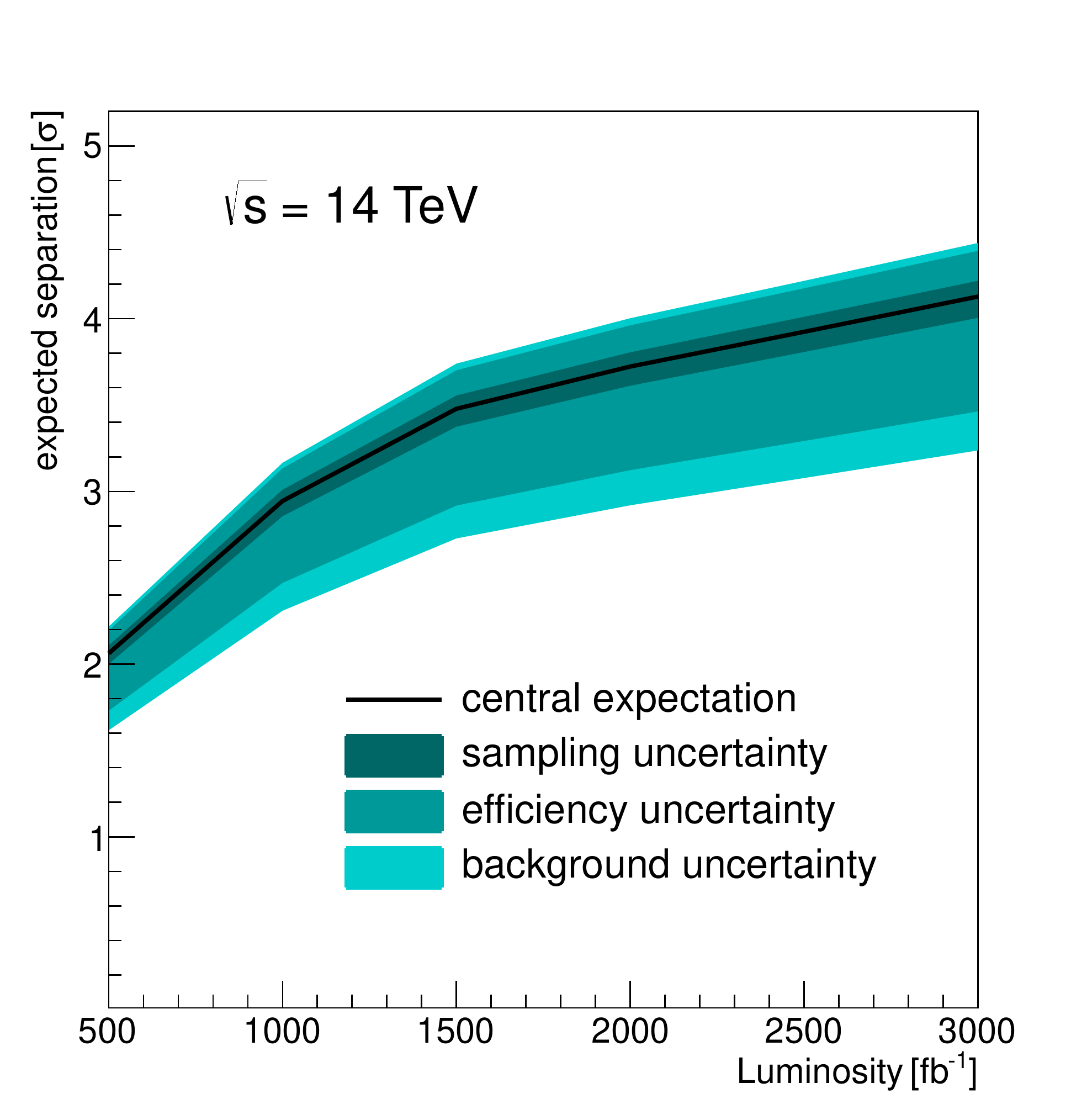}
    \includegraphics[width=\fwidth]{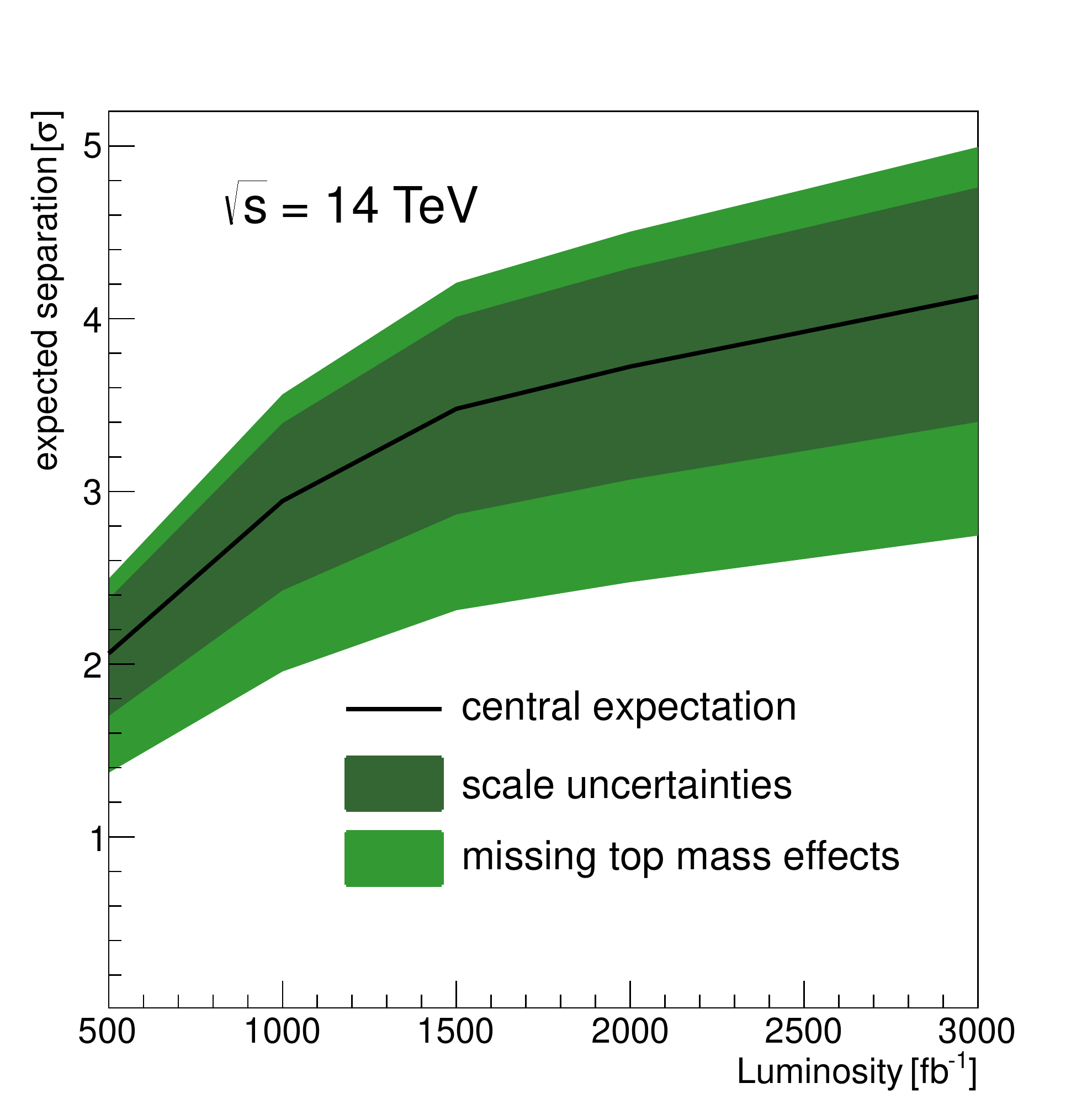}
    \caption{Expected sensitivity vs.~integrated luminosity. The
      experimental uncertainties (left) are combined quadratically,
      the theoretical uncertainties (right) are added linearly. }
    \label{result-lumi}
  \end{centering}
\end{figure}
% \clearpage

The two Higgs boson hypotheses can be separated with 2 standard
deviations ($\sigma$) with an integrated luminosity of about
$500\invfb$. With a luminosity of $1000\invfb$ the expected separation
sensitivity is about $3 \sigma$. The largest experimental uncertainty
(downwards) arises from the background uncertainty in the high-\pt\
region. A combination with other Higgs boson final states and a more
elaborate analysis would help to improve the sensitivity.  However,
theoretical uncertainties and missing top mass effects also have a
large impact on the sensitivity as can be inferred from
Fig.~\ref{result-lumi} (right panel). A reduction requires the full
implementation of higher-order effects on the transverse-momentum
distribution in the generators and in particular the full top mass
effects at NLO that have not been calculated so far.

%% file: conclusions.tex
\section{Conclusions}
\label{s:conclusions}
We have studied the possibility to separate in gluon fusion
loop-induced Higgs boson production from point-like production. Using
the Higgs boson yields (normalized to the overall rate) and the shape
of the Higgs boson \pt\ distribution the two hypotheses can be
separated with $2 \sigma$ with an integrated luminosity of
about $500\invfb$ (likely after Run-2 of the LHC). The largest
systematic uncertainty affecting this estimate is the background event
yield, as that strongly dilutes the difference between the two Higgs
boson hypotheses. Understanding and mitigating the impact from pile up
will be crucial for this analysis.

%% file: acknowledgements.tex
\section{Acknowledgements}
\label{s:acknowledgements}
We are grateful to S.~Frixione for his help with \mcatnlo\ and to
A.~Vicini for his help with \powheg. We thank our CMS colleagues, in
particular M.~Donega, for useful discussions.